\documentclass[letterpaper]{article} 
\usepackage{aaai2026}  
\usepackage{times}  
\usepackage{helvet}  
\usepackage{courier}  
\usepackage[hyphens]{url}  
\usepackage{graphicx} 
\usepackage{natbib}  
\usepackage{caption} 
\usepackage{algorithm}
\usepackage{algorithmic}

\usepackage{newfloat}
\usepackage{listings}
\DeclareCaptionStyle{ruled}{labelfont=normalfont,labelsep=colon,strut=off} 
\floatstyle{ruled}
\newfloat{listing}{tb}{lst}{}
\floatname{listing}{Listing}
\title{STEPWISE-CODEX-Bench: Evaluating Complex Multi-Function Comprehension and Fine-Grained Execution Reasoning}
\author{
    \textbf{Kaiwen Yan},
    \textbf{Yuhang Chang},
    \textbf{Zirui Guo},
    Yaling Mou,
    Jiang Ming,
    Jingwei Sun \textsuperscript{\dag}
}
\affiliations{

    \textbf{ByteDance}
}

\usepackage{bibentry}
\usepackage{bibentry}
\usepackage{booktabs}
\usepackage{multirow}
\usepackage{enumitem}  
\usepackage{booktabs}  
\usepackage{threeparttable}  
 \usepackage{amsmath} 

\begin{document}

\maketitle

\let\oldthefootnote\thefootnote 
\renewcommand{\thefootnote}{} 
\footnotetext{\textsuperscript{\dag}Project leader}
\let\thefootnote\oldthefootnote 
\begin{abstract}
In recent years, large language models (LLMs) have achieved notable breakthroughs in the field of code intelligence, yet in-depth evaluation of their code understanding and reasoning capabilities remains challenging. Existing mainstream benchmarks (e.g., HumanEval, MBPP) focus on the functional correctness of code generation, while reasoning benchmarks such as CRUXEVAL are limited to single-function, low-complexity scenarios. As a result, advanced current models generally score more than 95\% on these benchmarks, losing their discriminative power. To address this, this paper proposes STEPWISE-CODEX-Bench (SX-Bench), a new benchmark for complex multi-function understanding and fine-grained execution reasoning. Its core innovations include: constructing complex scenarios involving collaboration of multiple sub-functions (such as chained calls and nested loops), shifting the evaluation focus to the modeling of the overall control flow and data flow of programs; defining the minimum execution unit of a program as ``computation steps", and requiring models to predict the total number of steps in reasoning tasks to assess their in-depth understanding of dynamic execution trajectories, going beyond the superficial reasoning of traditional I/O matching.Evaluations of over 20 mainstream models (including 14 reasoning-enhanced models) show that this benchmark has strong discriminative power — even the current state-of-the-art model openai-o3 achieves an accuracy rate of only 78.37\% on Hard-Reasoning tasks, which is far lower than the saturated scores of existing benchmarks, effectively revealing the bottlenecks of existing models in complex logic and fine-grained reasoning. Meanwhile, we design and open source an automated task generation and quality assurance pipeline that integrates program synthesis, symbolic execution, and LLM-aided verification, providing an efficient path for building complex benchmarks. SX-Bench advances the evaluation of code understanding from ``single-function functional verification" to ``multi-function dynamic execution reasoning", offering a key tool for the in-depth capability evaluation of advanced code intelligence models.
\end{abstract}

\section{Introduction}

 In recent years, Large Language Models (LLMs) have demonstrated remarkable potential in the code intelligence domain, achieving significant breakthroughs, particularly in \textbf{ code generation, completion, and automation}~\cite{chai2024mcevalmassivelymultilingualcode,liu2024fullstackbenchevaluatingllms}. As model capabilities advance rapidly, accurately evaluating the depth and limitations of their \textbf{code comprehension and reasoning abilities } has become critical. However, current mainstream code generation benchmarks (e.g., HumanEval~\cite{chen2021evaluatinglargelanguagemodels}, MBPP~\cite{austin2021programsynthesislargelanguage}) primarily focus on whether models can generate functionally correct code snippets, with an emphasis on Input-Output (I/O) matching. Yet they overlook models’ ability to comprehension and reason about the internal execution processes of complex program logic. A truly powerful code intelligence model must not only \texttt{"write"} code, but also \texttt{"comprehension"} and \texttt{"reasoning"}the execution paths and internal state changes of complex code. 

Recent code comprehension and reasoning benchmarks like CRUXEVAL~\cite{gu2024cruxevalbenchmarkcodereasoning} and CRUXEVAL-X~\cite{xu2025cruxevalxbenchmarkmultilingualcode} attempt to address this by predicting function outputs/inputs, but suffer from critical limitations: (1) \textbf{Narrow scope}: Limited to single-function logic (even CRUXEVAL-X, despite multi-language support); (2) \textbf{Low complexity}: Evaluated functions are short (less 20 lines) with trivial logic; (3) \textbf{Loss of discriminability}: As shown in Table \ref{tab:crux_model_performance}, our evaluation shows that the SOTA models (Gemini-2.5-pro, openai-o3, etc.) achieve  exceed \textbf{95\%} precision in CRUXEVAL / X, failing to distinguish the capabilities of the model. These gaps indicate current benchmarks lag behind model development, necessitating more challenging evaluation frameworks.  

To bridge this gap, we present STEPWISE-CODEX-Bench \textbf{(SX-Bench)}, a novel benchmark for \textbf{complex multi-function comprehension} and \textbf{fine-grained execution reasoning}. Our core innovations are:  
\textbf{(1) Complex Multi-Function Scenarios}: Abandoning single-function constraints, SX-Bench designs programs with multi-subfunction collaboration ($\geq 3$ sub-functions) and complex interactions (chained calls, nested loops, multi-branch conditionals), shifting evaluation from single-point mapping to control/data flow modeling.  
\textbf{(2) Stepwise Execution Tracing Paradigm}:  
We define each \textbf{atomic computation unit} in a program as a ``computation step." For example, complex processes such as loop iterations, conditional branch evaluations, and cross-function calls typically consist of multiple computation steps, whereas basic operations in arithmetic/logical operation (e.g. addition, subtraction, multiplication, division) represent the minimal granularity of computation units. To precisely quantify this process, we embed a global counter variable \texttt{run\_steps} in the program. Each time the program executes one of the above-defined basic operations, the value of \texttt{run\_steps} is automatically incremented by 1. After the program finishes execution, the final value of \texttt{run\_steps} represents the total number of computation steps that the program has undergone from start to finish. In evaluation tasks, we require the model to predict the value of \texttt{run\_steps} after program execution based on the given program code and input.
We evaluated \textbf{20+ mainstream models} (14 reasoning enhanced LLMs) on SX-Bench. Results show strong discriminability: even top model \textbf{openai-o3} achieves only \textbf{78.37\% accuracy} on Hard-Reasoning, highlighting SX-Bench’s ability to capture reasoning bottlenecks.  

To support SX-Bench’s scalability, we developed a automated task generation pipeline integrating program synthesis, symbolic execution, and LLM-assisted validation, reducing costs for high-quality, complex samples.  

In summary, our contributions are:  
\begin{enumerate}  
    \item Proposing \textbf{SX-Bench}, a pioneering benchmark for complex multi-function comprehension and fine-grained execution reasoning.  
    \item Conducting comprehensive evaluations of over 20 mainstream LLMs (including 14 reasoning enhanced models) on SX-Bench, revealing significant bottlenecks in existing models' deep code understanding   
    \item Designing and open-sourcing an automated code comprehension task generation and quality assurance pipeline, providing a practical pathway for future complex benchmark construction.  
\end{enumerate}

\begin{table}
  \centering
  \begin{tabular}{lcc}  
    \toprule  
    Model               & CRUXEVAL & CRUXEVAL-X \\  
    \midrule  
    Gemini-2.5-Pro      & 98.2     & 97.8       \\  
    openai-o4-mini         & 98.0     & 98.5       \\  
    Deepseek-R1         & 97.6     & 96.7       \\  
    Doubao-Seed-1.6     & 95.3     & 95.9       \\  
    \bottomrule  
  \end{tabular}
  \caption{Performance of SOTA Models on CRUXEVAL and CRUXEVAL-X Benchmarks} 
  \label{tab:crux_model_performance}
\end{table}
\section{SX-Bench}
\subsection{Overview}
\begin{figure*}
    \centering
    \includegraphics[width=\textwidth]{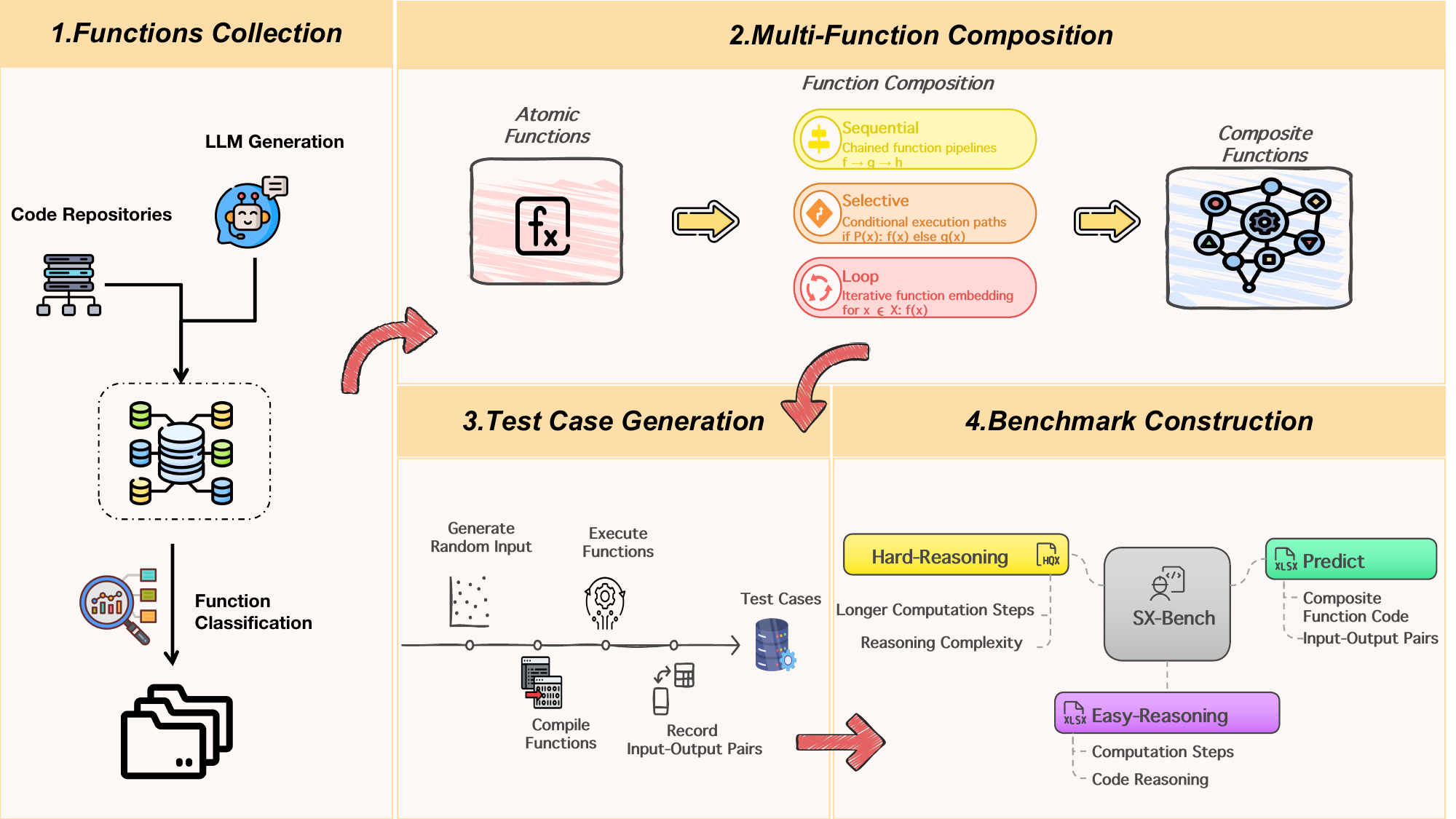}
    \caption{The construction Pipeline of SX-Bench}
    \label{fig:main_pipeline}
\end{figure*}

\textbf{SX-Bench} is a new benchmark suite designed for complex \textbf{code comprehension }and \textbf{fine-grained execution reasoning}, aimed at addressing the limitations of existing code comprehension evaluation frameworks in assessing deep program analysis capabilities. Its core design philosophy shifts from \texttt{"superficial comprehension of static code"} to \texttt{"modeling dynamic execution processes"}, systematically evaluating the reasoning capabilities of LLMs regarding multifunction collaboration logic, control flow/data flow interactions, and fine-grained execution steps. The overall construction process is as shown in Figure \ref{fig:main_pipeline}.

\subsection{Data Construction}
\paragraph{Collection and Classification of Single Functions}  
To construct a library of \textbf{basic function units}, we first define the fundamental specifications for \textbf{single functions}: functions with identical input-output mapping relationships are classified into the same class. Based on this criterion, we collected a total of \textbf{30 function classes} containing \textbf{361 specific function instances} in both Go and Python, through \textbf{LLM generation} and \textbf{basic code repositories}. These functions cover essential functionalities such as mathematical operations (e.g., arithmetic operations like addition, subtraction, multiplication, division, and exponentiation), logical judgments (e.g., parity check, magnitude comparison), and string manipulation (e.g., concatenation, substring extraction).


\paragraph{Multi-Function Composition Construction}
Building on the atomic function library, we construct \textbf{composite functions} by orchestrating execution patterns across multiple atomic functions. Our framework implements three fundamental composition paradigms:
\begin{description}[leftmargin=12mm, noitemsep, topsep=0pt]
\item[\textbf{Sequential:}] Output-to-input propagation across function pipelines ($f \circ g \circ h$)
\item[\textbf{Selective:}] Conditional execution paths ($\text{if } P(x): f(x) \text{ else } g(x)$)
\item[\textbf{Loop:}] Loop-embedded function iterations ($\text{for } x \in X: f(x)$)
\end{description}

\paragraph{Test Case Generation}  
To evaluate the behavior of \textbf{composite functions}, we designed an automated test case generation suite: for each composite function, \textbf{random input data} are generated. The composite functions are then compiled and executed through a compiler, with \textbf{input-output pairs} (i.e.,$<$input, actual output$>$) recorded.

\paragraph{Test Case Quality Inspection}  
Random inputs may trigger abnormal behaviors in composite functions (e.g., infinite loops, execution timeout with a threshold set to 3 seconds, numeric overflow where integers exceed the 64-bit range), rendering test cases invalid. To address this, we designed a \textbf{multistage filtering mechanism}:  

\begin{itemize}[leftmargin=*]  
  \item \textbf{Execution Monitoring}: Test cases are run in a sandbox environment to intercept infinite loops and timeout cases.  
  \item \textbf{Numeric Validation}: Detection of whether outputs exceed the predefined numeric range 
  \item \textbf{Logical Consistency}: The same input is executed multiple times to verify output stability.  
\end{itemize}  

Ultimately, valid test cases that pass quality inspection are retained.

\subsection{Benchmark Construction}

Existing code comprehension benchmarks, including CRUXEVAL, typically require models to directly output code execution results, but models may be misjudged due to insufficient instruction-following capabilities. To address this issue, we designed an evaluation paradigm based on ``input-output pair matching judgment": instead of generating results, models only need to determine whether a given $<$input, output$>$ pair is matched—that is, whether the output equals the result of executing the function on the input—and finally output ``Yes" or ``No" to eliminate the impact of format errors. Based on this, we constructed three benchmark subsets:

\begin{itemize}[leftmargin=*]
\item \textbf{Predict}: A code comprehension benchmark. The input includes composite function code and multiple $<$input, output$>$ pairs, with the task of judging whether each pair is matched.
\item \textbf{Easy-Reasoning}: An easy code reasoning benchmark. We define ``computation steps" as the minimal execution unit of a program, The input includes composite function code and specific inputs, with the task of outputting the total number of computation steps corresponding to the input.
\item \textbf{Hard-Reasoning}: A hard code reasoning benchmark. Compared to the easy version, it contains more single-function combinations, more functions, and longer computation steps, with step counts four times that of the easy version, enhancing reasoning complexity.
\end{itemize}

To prevent models from scoring via random guessing, all benchmark samples satisfy:
\begin{itemize}[leftmargin=*]
\item Each sample contains at least 3 $<$input, output$>$ pairs;
\item 50\% of the generated input-output pairs are randomly selected to be ``corrupted" (i.e., the output is modified to mismatch the input), ensuring the match/mismatch ratio is approximately 1:1.
\end{itemize}

\begin{table*}[t!]
\centering
\begin{tabular}{@{}lccccccc@{}} 
\toprule
\textbf{Subset} & 
\multicolumn{1}{c}{\textbf{Total Samples}} & 
\multicolumn{1}{c}{\textbf{Avg. Tests}} & 
\multicolumn{1}{c}{\textbf{Avg. Funcs}} & 
\multicolumn{1}{c}{\textbf{Avg. Steps}} & 
\multicolumn{1}{c}{\textbf{Sequential}} & 
\multicolumn{1}{c}{\textbf{Loop}} & 
\multicolumn{1}{c}{\textbf{Selective}} \\
\midrule
Predict & 626 & 7.20 & 8.95 & -- & 208 & 204 & 214 \\
Easy-Reasoning & 750 & 7.83 & 8.24 & 14.92 & 325 & 184 & 241 \\
Hard-Reasoning & 504 & 9.00 & 9.10 & 69.49 & 116 & 216 & 172 \\
\bottomrule
\end{tabular}
\caption{Statistics of SX-Bench Evaluation Subsets. 
Note: Avg. Tests = Average number of input-output pairs per sample; 
Avg. Funcs = Average number of functions per composition; 
Avg. Steps = Mean computation steps per input.}
\label{tab:benchmark-stats}
\end{table*}

\subsection{Data Analysis}
To validate the design objectives of SX-Bench, we performed statistical analyzes of the key characteristics of the three subsets, the results shown in Table \ref{tab:benchmark-stats}.

\paragraph{Sample Scale and Coverage}
SX-Bench comprises a total of 1,880 high-quality samples, including 626 in the Predict subset, 750 in the Easy-Reasoning subset, and 504 in the Hard-Reasoning subset. The allocation of sample sizes across subsets not only ensures completeness in evaluating basic comprehension abilities, but also maintains the challenge of advanced reasoning tasks. In terms of test cases, each sample contains an average of \textbf{7.20} to \textbf{9.00} input-output pairs. By randomly ``corrupting" 50\% of the output values, we constructed a balanced binary classification task to ensure that the evaluation results truly reflect the model's understanding.

In particular, the composite functions in each subset contain an average of \textbf{8.24} to \textbf{9.10} subfunctions, far exceeding single function benchmarks such as CRUXEVAL. The Hard-Reasoning subset features a higher number of functions and more complex sub-function interactions, further enhancing the requirements for models to track data flows.

\paragraph{Distribution of Reasoning steps}
\begin{figure}
    \centering
    \includegraphics[width=0.48\textwidth]{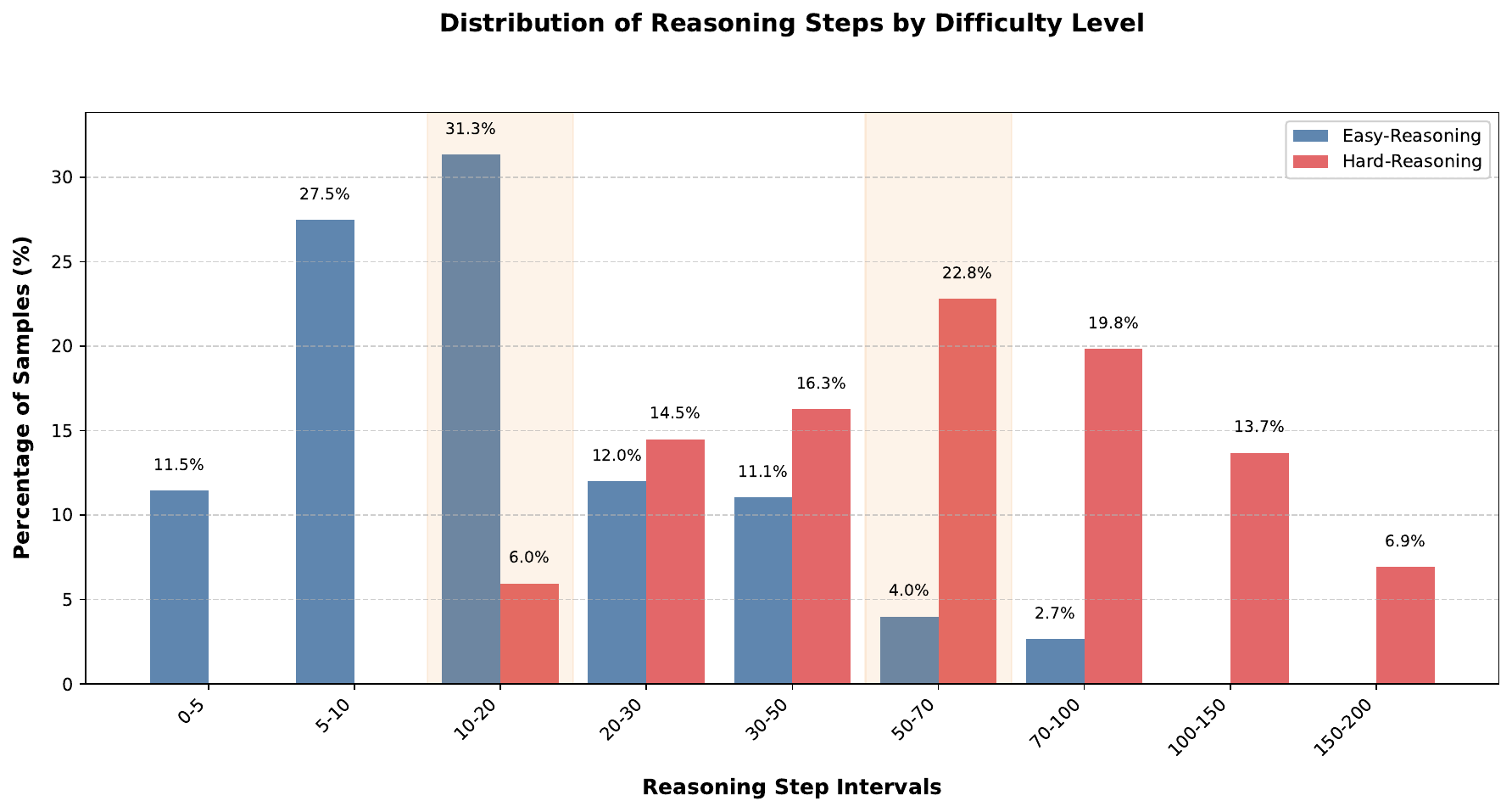}
    \caption{ The distribution of reasoning steps in datasets of different difficulties.}
    \label{fig:reasoning_depth_distribution}
\end{figure}
\texttt{\textbf{"Computation step reasoning"}} serves as the core task of \textbf{SX-Bench} for evaluating fine-grained execution reasoning, with two datasets designed at different difficulty gradients. The distribution of reasoning depth across datasets is shown in Figure \ref{fig:reasoning_depth_distribution}. The \textbf{Easy-Reasoning} subset has an average step count of \textbf{14.92}, corresponding to composite functions with simple loops or single conditional branches, evaluating models' basic execution tracing capabilities. The \textbf{Hard-Reasoning} subset reaches an average of \textbf{69.49} steps—approximately \textbf{4.65} times that of the Easy subset—primarily driven by complex structures and frequent function calls, requiring models to precisely handle fine-grained reasoning tasks.

\paragraph{Diversity of Composition Pattern}
\begin{figure}
    \centering
    \includegraphics[width=.48\textwidth]{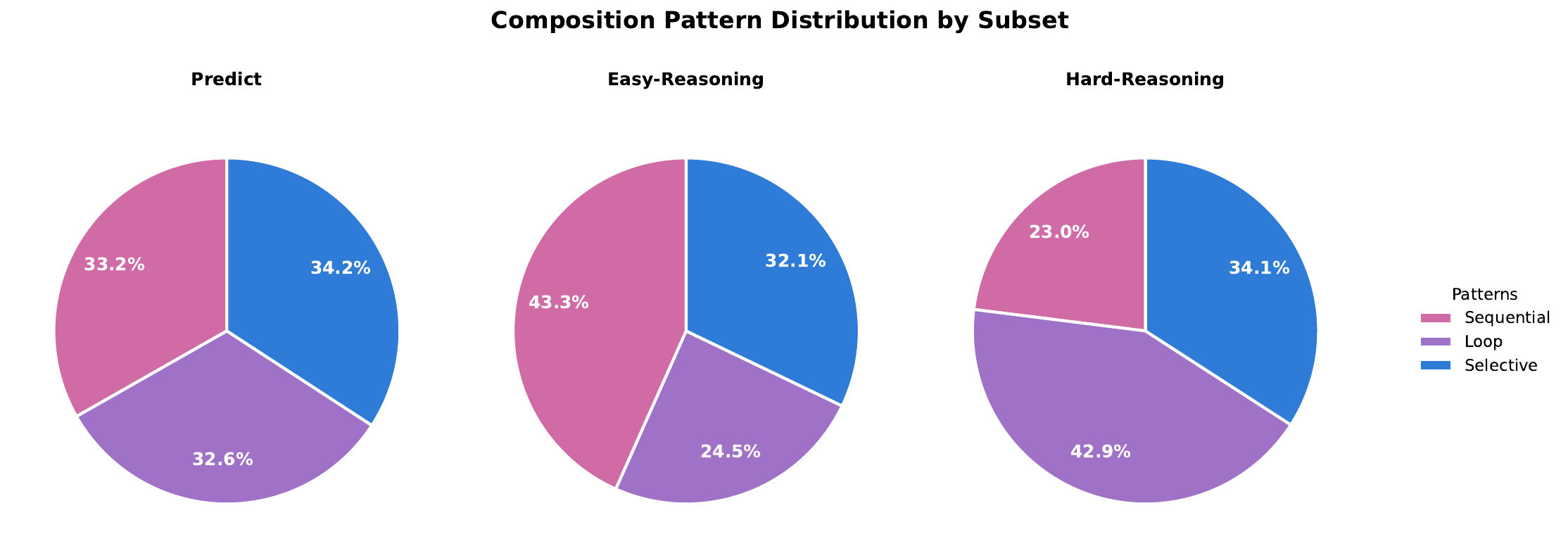}
    \caption{ The distribution of Composition Pattern Distribution by Subset}
    \label{fig:benchmark_patterns}
\end{figure}
\textbf{SX-Bench} constructs composite functions through three fundamental composition paradigms: \texttt{\textbf{"Sequential"}}, \texttt{\textbf{"Loop"}}, and \texttt{\textbf{"Selective"}}, featuring balanced sample distribution and clear evaluation focuses for each paradigm. In the Predict subset, various composition paradigms are evenly distributed. Sequential chaining demonstrates the highest proportion in the Easy-Reasoning subset, examining the model's understanding of data flow transmission; the proportion of loop iteration significantly increases in the Hard-Reasoning subset, focusing on evaluating the model's counting ability for iterative processes; selective invocation maintains balanced distribution across all subsets. The distribution ratios are illustrated in Figure \ref{fig:benchmark_patterns}.

\section{Experiment}

\subsection{Experiment setting}
\paragraph{Evaluated Models}
We selected a total of 29 representative models across two categories, spanning different vendors, parameter scales, and reasoning/non-reasoning capabilities:
\begin{itemize}
\item \textbf{Non-Reasoning Models}: including Qwen2.5 Series\cite{qwen2025qwen25technicalreport} (32B/14B/72B), Llama-3.3-70B\cite{grattafiori2024llama3herdmodels}, Qwen3 Series\cite{yang2025qwen3technicalreport} (14B/32B/30A3B/235A22B), DeepSeek-V3\cite{deepseekai2025deepseekv3technicalreport}, Doubao-1.5-Pro\cite{seed2025seed15thinkingadvancingsuperbreasoning}, GLM-4-Air\cite{glm2024chatglmfamilylargelanguage},Baichuan-4-Turbo,and GPT-4.1.
\item \textbf{Reasoning-Enhanced Models}:Models that handle complex tasks and generate accurate and logically consistent results through systematic thinking such as step-by-step decomposition, logical deduction, and verification, and enhance reasoning capabilities via Chain-of-Thought prompting (e.g., guidance with ``Think"). These include reasoning modes of Qwen3 Series\cite{yang2025qwen3technicalreport} (32B/30A3B/235A22B), Qwen-QwQ-32B\cite{qwq32b}, MiniMax-M1\cite{minimax2025minimaxm1scalingtesttimecompute},GLM-Z1-Air\cite{glm2024chatglmfamilylargelanguage} Doubao-1.5-thinking-pro\cite{seed2025seed15thinkingadvancingsuperbreasoning}, Doubao-1.6, DeepSeek-R1\cite{deepseekai2025deepseekr1incentivizingreasoningcapability}, Gemini-2.5-flash/pro\cite{gemini2.5}, openai-o4-mini, and openai-o3.
\end{itemize}

\paragraph{Evaluation Metric}
We adopt Accuracy as the core metric, defined as follows: A sample contains multiple independent judgment tasks (e.g., the Predict subset has an average of 7.20 sets of $<$input, output$>$ pairs to judge). The model must correctly complete all tasks within the sample for the sample to be considered correct, so as to rigorously evaluate its reasoning consistency.

\paragraph{Experimental Setup}
Evaluation environment configuration is as follows: We use a computing cluster of 8 NVIDIA A800 (40G) GPUs as the evaluation device. For the inference framework, non-API models are locally deployed via the \textbf{vllm}\cite{kwon2023efficient} framework, while API models directly call official interfaces such as OpenAI API and Gemini API. Key parameters are uniformly set for all models: max\_tokens is set to \textbf{16K} to ensure complete output of long reasoning processes, temperature is set to \textbf{0} to suppress randomness and guarantee reproducible results, and input formats strictly follow the official instruction specifications of the models.

\begin{table*}[!ht]
\centering
\resizebox{1.2\columnwidth}{!}{
\begin{tabular}{lcccc}  
\toprule
\multirow{2}{*}{Models} & \multicolumn{4}{c}{\textbf{Non-Reasoning Models}}\\
\cmidrule(lr){2-5} 
&Overall & Predict & Easy-Reasoning & Hard-Reasoning  \\
\midrule
Qwen2.5-Coder-14B-Instruct	&8.90&	18.85&	6.27&	1.59\\
Qwen2.5-Coder-32B-Instruct & 12.12 & 22.84 & 11.33 & 2.18 \\
Qwen2.5-14B-Instruct & 8.68 & 18.85 & 6.0 & 1.19 \\
Qwen2.5-32B-Instruct & 13.12 & 24.44 & 12.53 & 2.38 \\
Qwen2.5-72B-Instruct & 14.49 &	27.16	&13.33&	2.98 \\
GLM-4-Air&	4.94&	9.74	&4.67&	0.40\\
Baichuan-4-Turbo&	5.59	&9.58&	6.00&	1.19\\
Llama-3.3-70b-Instruct & 10.89 & 21.41 & 9.47 & 1.79 \\
Qwen3-14B & 12.78 & 21.25 & 13.33 & 3.77 \\
Qwen3-32B & 16.68 & 29.07 & 16.4 & 4.56 \\
Qwen3-30A3B & 16.40 & 28.91 & 16.93 & 3.37 \\
Qwen3-235A22B & 18.10 & 27.80 & 21.73 & 4.76 \\
DeepSeek-V3 & 22.30 & 37.22 & 22.53 & 7.14 \\
Doubao1.5-Pro & 19.64 & 31.15 & 22.4 & 5.36 \\
GPT-4.1 & \textbf{36.12} & \textbf{52.88} & \textbf{40.80} & \textbf{14.68} \\
\midrule
\multirow{2}{*}{Models} & \multicolumn{4}{c}{\textbf{Reasoning-Enhanced Models}}\\
\cmidrule(lr){2-5} 
&Overall & Predict & Easy-Reasoning & Hard-Reasoning  \\
\midrule
Qwen3-14B(Think) &	45.91&	71.09&	47.2	&19.44\\
Qwen3-32B(Think) & 55.93 & 74.6 & 57.87 & 35.32 \\
Qwen3-30A3B(Think) & 44.26 & 67.73 & 45.6 & 19.44 \\
Qwen3-235A22B(Think) & 54.03 & 73.80 & 58.93 & 29.37 \\
Qwen-QwQ-32B & 57.87 & 80.67 & 59.2 & 33.73 \\
GLM-Z1-Air &46.94&	68.69&	46.93&	25.2\\
MiniMax-M1 & 48.62 & 76.52 & 47.33 & 22.02 \\
Doubao-1.5-thinking-pro & 72.66 & 89.14 & 75.07 & 53.77 \\
Doubao-Seed-1.6(Think) & 76.13 & 87.86 & 81.6 & 58.93 \\
DeepSeek-R1 & 73.54 & 84.98 & 81.07 & 54.56 \\
Gemini-2.5-flash & 54.31 & 65.66 & 66.13 & 31.15 \\
Gemini-2.5-Pro & 82.89	&90.10&	90.13&	68.45\\
openai-o4-mini & 82.27 & 90.42 & 87.73 & 68.65 \\
openai-o3 & \textbf{86.92} & \textbf{91.05} & \textbf{91.33} & \textbf{78.37} \\
\bottomrule
\end{tabular}
}
\caption{Performance Comparison of Non-Reasoning and Reasoning-Enhanced Models on SX-Bench}
\label{tab:model_evaluation_results}
\end{table*}

\subsection{Main Experiments}

\paragraph{Overall Performance Comparison}

Table\ref{tab:model_evaluation_results} presents the overall performance of non-reasoning models and reasoning-enhanced models on SX-Bench. Reasoning ability is critical for deep code understanding: reasoning-enhanced models achieved an average performance of \textbf{64.5\%}, significantly outperforming non-reasoning models (\textbf{17.8\%} on average) with an improvement of nearly \textbf{3.6} times. Even the top-performing non-reasoning model, \textbf{GPT-4.1}, still lagged far behind the worst-performing reasoning-enhanced model, indicating that code execution reasoning requires dedicated reasoning mechanisms.

SX-Bench exhibits strong discriminative power: The state-of-the-art reasoning-enhanced model, \textbf{openai-o3}, achieved an overall accuracy of \textbf{86.92\%}, without the ``ceiling effect" (where SOTA models exceed 95\%) observed in existing benchmarks like CRUXEVAL. Notably, its accuracy on the Hard-Reasoning subset was only \textbf{78.37\%}, validating SX-Bench's effective discrimination of complex reasoning abilities.

\paragraph{Performance Analysis by Subset}
Subset-level analysis reveals that SX-Bench’s three progressive evaluation subsets exhibit distinct difficulty gradients and model capability discrimination. In SX-Bench-Predict, which evaluates judging input-output matching of composite functions, reasoning-enhanced models achieved over 90\% accuracy, while non-reasoning models (e.g., \textbf{GPT-4.1}) reached a maximum of only \textbf{52.88\%}, highlighting the limitations of non-reasoning models in understanding dependencies of multi-function control flow and data flow.

In Easy-Reasoning, reasoning-enhanced models maintained over 90\% accuracy, effectively handling single-loop and simple conditional branch counting. However, in Hard-Reasoning—featuring complex nested loops, high-frequency function calls, and long execution paths—model performance dropped significantly. Even the model with the highest performance of reasoning enhanced, \textbf{openai-o3}, achieved only \textbf{78.37\%} accuracy, a 13-percentage-point decrease from the Easy subset. Additionally, the performance gap among reasoning-enhanced models widened (e.g. \textbf{Gemini-2.5-Pro} at \textbf{67.30\%} vs. \textbf{openai-o3} at \textbf{78.37\%}, a 11-percentage-point difference), while non-reasoning models almost failed (with a maximum accuracy of only \textbf{7.14\%}). These results confirm that modeling complex dynamic execution processes remains a core bottleneck for current models.
\section{In-Depth Analysis}

\subsection{Model Capability Analysis}
\begin{figure*}[!ht]
    \centering
    \includegraphics[width=0.8\textwidth]{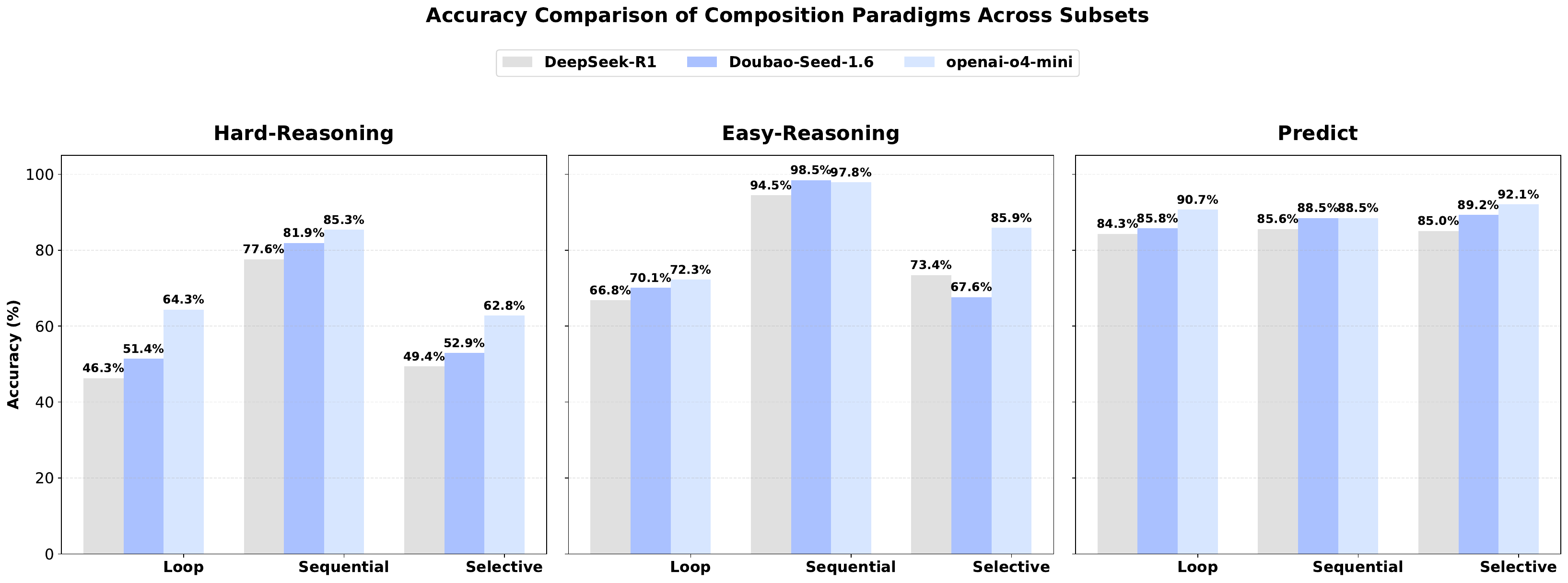}
    \caption{ The Accuracy Comparison of Composition Paradigms Across Subsets.}
    \label{fig:accuracy_comparison}
\end{figure*}
\paragraph{Performance Analysis Across Composition Paradigms}
We selected three models \textbf{DeepSeek-R1}, \textbf{Doubao-Seed-1.6}, and \textbf{openai-o4-mini} to analyze the impact of difficulty across different atomic composition paradigms. As shown in Figure \ref{fig:accuracy_comparison}, the three models exhibit significant performance differences across different composition paradigms (\textbf{Loop}, \textbf{Sequential}, \textbf{Selective}) and subsets; overall, the difficulty distribution is \textbf{Loop} $>$ \textbf{Selective} $>$ \textbf{Sequential}. Across all subsets, all models consistently achieve the highest accuracy in the \textbf{Sequential} paradigm. This is likely because the \textbf{Sequential} paradigm involves a linear, step-by-step reasoning process with clear logical chains, which models can process more straightforwardly. In contrast, \textbf{Loop} and \textbf{Selective} paradigms exhibit lower accuracy, likely due to their higher complexity, the \textbf{Loop} paradigm requires iterative or conditional reasoning, while the \textbf{Selective} paradigm demands that models make decisions on which subtasks to execute, both of which introduce additional cognitive load.

\begin{table}[h]
\centering
\resizebox{0.45\textwidth}{!}{
\begin{tabular}{lccccc}
\toprule
\textbf{Model} & \textbf{Type} & \textbf{Overall} & \textbf{Predict} & \textbf{Easy-Reasoning} & \textbf{Hard-Reasoning} \\
\midrule
\multirow{2}{*}{Qwen3-14B} & Non-Reasoning & 12.78 & 21.25 & 13.33 & 3.77 \\
& Reasoning-Enhanced & 45.91 & 71.09 & 47.2 & 19.44  \\
\midrule
\multirow{2}{*}{Qwen3-32B} & Non-Reasoning & 16.68 & 29.07 & 16.4 & 4.56  \\
& Reasoning-Enhanced & 55.93 & 74.6 & 57.87 & 35.32  \\
\bottomrule
\end{tabular}}
\caption{Non-Reasoning vs. Reasoning Performance Comparison}
\label{tab:reasoning_cap}
\end{table}

\paragraph{Reasoning vs. Non-Reasoning Mode Performance Comparison}
As shown in Table \ref{tab:reasoning_cap}, we selected \textbf{Qwen3-14B} and \textbf{Qwen3-32B} for comparison. For both models, enabling the reasoning-enhanced mode led to significant improvements in accuracy across all subsets, particularly in the \textbf{Hard-Reasoning subset}. This indicates that reasoning capabilities play a critical role in task performance.

Specifically, in terms of overall performance: \textbf{Qwen3-14B} increased from \textbf{12.78} in non-reasoning mode to \textbf{45.91}, with an increase of \textbf{33.13}, representing approximately \textbf{259\%} improvement; \textbf{Qwen3-32B} increased from \textbf{16.68} to \textbf{55.93}, with an increase of \textbf{39.25} and an approximate \textbf{235\%} improvement.

In non-reasoning mode, \textbf{Qwen3-32B} scored \textbf{3.90} points higher overall than \textbf{Qwen3-14B}, with gaps of \textbf{2.73} in Easy-Reasoning and \textbf{0.79} in Hard-Reasoning. In reasoning-enhanced mode, the overall score gap increased to \textbf{ 10.02}, including a gap of \textbf{15.88} in Hard-Reasoning, demonstrating that larger models can better leverage reasoning mechanisms to handle complex tasks.

\begin{figure}
    \centering
    \includegraphics[width=0.45\textwidth]{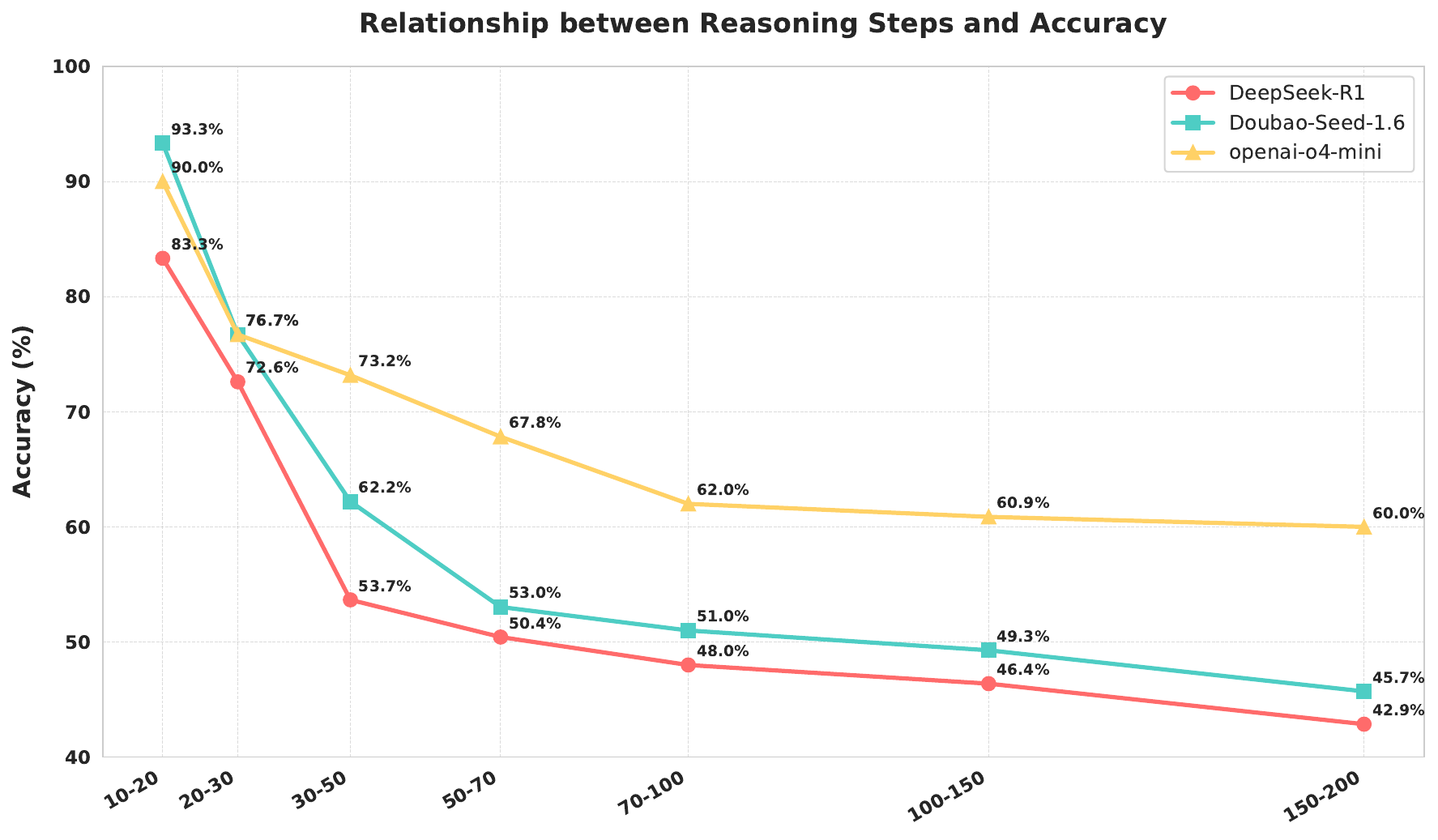}
    \caption{ The Accuracy Comparison of Different Reasoning Steps}
    \label{fig:reasoning_depth_vs_accuracy}
\end{figure}

\paragraph{Analysis of the Performance of Different Reasoning Steps}
As shown in Figure \ref{fig:reasoning_depth_vs_accuracy}, we analyzed the performance of three models \textbf{DeepSeek-R1}, \textbf{Doubao-Seed-1.6}, and \textbf{openai-o4-mini} on data requiring different numbers of reasoning steps within the \textbf{Hard-Reasoning} subset.

Across all three models, accuracy in the \textbf{Hard-Reasoning} subset consistently declined as the number of reasoning steps increased. This trend indicates that longer reasoning chains significantly increase task difficulty, requiring stronger capabilities in maintaining logical coherence, mitigating error accumulation, and modeling complex relationships. \textbf{openai-o4-mini} demonstrated greater robustness in high-step scenarios, likely due to its superior context retention and reasoning chain correction mechanisms. \textbf{Doubao-Seed-1.6} excelled in straightforward low-step reasoning tasks but exhibited rapid performance degradation beyond 100 steps, suggesting room for improvement in verification of multistep logical and intermediate results. \textbf{DeepSeek-R1} showed the weakest overall reasoning ability, possibly constrained by inefficient planning and execution of complex reasoning paths. These findings reinforce the Hard-Reasoning subset's role as a rigorous test of deep reasoning capabilities, highlighting that the cumulative cognitive load from increased steps is a critical performance bottleneck.

\subsection{Enhance Model Code Comprehension and reasoning capabilities}
Based on the foregoing analysis, the performance of models in code reasoning tasks is jointly influenced by reasoning modes, model scale, and task complexity. To enhance reasoning capabilities in complex code scenarios, targeted improvements can be made in the following aspects:

\paragraph{Reasoning Mode Enhancement}
Reasoning mechanisms are the core engine for breaking through the bottleneck of code reasoning. Experimental data show that reasoning-enhanced models achieved an average accuracy of \textbf{64.5\%} on SX-Bench, significantly outperforming non-reasoning models, which had an average performance of \textbf{17.8\%} (Table \ref{tab:model_evaluation_results}).

After\textbf{ Qwen3-14B} and \textbf{Qwen3-32B} enabled the reasoning mode, the accuracy of the Hard-Reasoning subset increased from \textbf{3.77\%} to \textbf{19.44\%} and from \textbf{4.56\%} to \textbf{35.32\%} respectively, with an increase of \textbf{416\%} and \textbf{674\%} (Table \ref{tab:reasoning_cap}), indicating that the reasoning mode can effectively activate the model's ability to model complex logical chains.

\paragraph{Model Scale and Reasoning Mechanisms}
Model scale and reasoning mechanisms show significant synergy: larger scales mean more pronounced gains from reasoning modes. For instance, the Hard-Reasoning accuracy of \textbf{Qwen3-32B} in reasoning mode (\textbf{35.32\%}) is \textbf{15.88} percentage points higher than that of \textbf{Qwen3-14B} (\textbf{19.44\%}), where as the gap between the two in non-reasoning mode is only \textbf{0.79} percentage points (Table \ref{tab:reasoning_cap}). Large-scale models can further unleash their reasoning potential by enhancing long-context reasoning capabilities and increasing pretraining on ultra-long code execution trajectories; medium and small models can compensate for their own scale limitations by transferring the reasoning chains of large models through ``reasoning knowledge distillation."

\paragraph{Structured Code Understanding}
The model's insufficient understanding of complex structural paradigms is a key cause of the decline in \textbf{Hard-Reasoning} performance. Experiments show that the difficulty distribution of different structural paradigms is Loop $>$ Selective $>$ Sequential; even the best-performing reasoning model, openai-o3, still has significantly lower accuracy in the Loop paradigm than in the Sequential paradigm (Figure \ref{fig:accuracy_comparison}). Moreover, the \textbf{Hard-Reasoning} subset, which contains complex structures such as nested loops and high-frequency function calls, leads to an average 13-percentage point drop in model performance compared to the \textbf{Easy-Reasoning} subset. 

To address this, structural information from Abstract Syntax Trees (\textbf{AST}) and Control Flow Graphs (\textbf{CFG}) can be incorporated during the pretraining phase to strengthen the model's explicit modeling of the iterative logic of Loop structures and the conditional branch decision-making of Selective structures. In the reasoning phase, a ``structure recognition - block reasoning" strategy can be adopted for complex code: for example, prioritize parsing the range of iterative variables in nested loops and the execution path dependencies of conditional branches, then verify intermediate results step by step to reduce error accumulation in long-chain reasoning.
\section{Related Work}
\subsection{Code Generation Evaluation}
\paragraph{Single-Function Generation Benchmarks}
HumanEval~\cite{chen2021evaluatinglargelanguagemodels} was the first widely adopted code generation benchmark, consisting of natural language descriptions and test cases for 164 Python functions. It requires models to generate complete function implementations. Based on this, MBPP~\cite{austin2021programsynthesislargelanguage} expanded the scope to 1000 problems and provided more comprehensive test cases. These benchmarks have propelled the development of code generation models such as CodeLlama~\cite{rozière2024codellamaopenfoundation}. 

\paragraph{Multilingual and Scenario Expansion}
MultiPL-E~\cite{cassano2022multiplescalableextensibleapproach} extended HumanEval to 10 programming languages, validating models' cross-lingual generalization capabilities; MCEval~\cite{chai2024mcevalmassivelymultilingualcode} further expanded coverage to 58 languages, focusing on code generation performance for low-resource languages. FullStackBench~\cite{liu2024fullstackbenchevaluatingllms} attempts to evaluate full stack development capabilities, covering scenarios such as front-end components and back-end APIs. Building on this, CodeIF~\cite{yan2025codeifbenchmarkinginstructionfollowingcapabilities} focuses on instruction-following capabilities in code generation. This work revealing differences in models' capabilities in parsing and executing complex instructions.

\subsection{Code Comprehension and Reasoning Benchmarks}
\paragraph{Single-Function Behavior Prediction}
CRUXEVAL~\cite{gu2024cruxevalbenchmarkcodereasoning} was the first benchmark dedicated to code reasoning, requiring models to predict the output of a given Python function for specific inputs or infer inputs from given outputs. Its extended version, CRUXEVAL-X~\cite{xu2025cruxevalxbenchmarkmultilingualcode}, supports 10+ languages but remains confined to single functions and short code snippets (average $<$20 lines).
\section{Conclusion}
To overcome the limitations of current benchmarks (single-function focus, low complexity, and poor discriminative power), we propose \textbf{STEPWISE-CODEX-Bench} (\textbf{SX-Bench}), a novel benchmark for evaluating fine-grained execution reasoning in complex multi-function scenarios. By introducing multi-function collaboration patterns and a ``stepwise execution tracing" paradigm, \textbf{SX-Bench} shifts evaluation from superficial I/O matching to dynamic execution process modeling. Experiments show strong discriminative power: even advanced models like \textbf{openai-o3} achieve only \textbf{78.37\%} accuracy on Hard-Reasoning tasks (vs. \textbf{$>$95\%} on existing benchmarks), revealing bottlenecks in complex logic. Further analysis confirms performance degradation with longer reasoning steps and weaknesses in loops/nested calls. \textbf{SX-Bench} drives a paradigm shift from ``single-function verification" to ``multi-function dynamic reasoning". With current support for multiple programming languages including Go and Python. Future work will further expand language coverage and incorporate real-world execution scenarios, pushing the boundaries of code intelligence research.

\bibliography{aaai2026}

\end{document}